\documentclass[12pt]{iopart}

\usepackage{iopams}
\usepackage{color}
\usepackage{graphicx}
\usepackage{units}
\usepackage{soul}
\usepackage{lineno,notes2bib}

\begin{document}

\title[Radio frequency  occupancy state control of a single nanowire quantum dot]{Radio frequency occupancy state control of a single nanowire quantum dot}
\author{Matthias Wei\ss$^{1,3}$,
Florian J. R. Sch\"ulein$^{1,3}$,
J\"{org} B. Kinzel$^{1,3}$,
Michael Heigl$^{1}$,
Daniel Rudolph$^{2,3}$, 
Max Bichler$^{2}$,
Gerhard Abstreiter $^{2,3,4}$,
Jonathan J. Finley$^{2,3}$,
Achim Wixforth$^{1,3,5}$,
Gregor Koblm\"{u}ller$^{2,3}$, 
Hubert J. Krenner$^{1,3,5,*}$}

\address{$^1$ Lehrstuhl f\"ur Experimentalphysik 1 and Augsburg Centre for Innovative Technologies {\it ACIT}, Universit\"{a}t Augsburg, Universit\"{a}tsstra\ss e 1, 86159 Augsburg, Germany}
\address{$^2$ Walter Schottky Institut and Physik Department, Technische Universit\"{a}t M\"{u}nchen, Am Coulombwall 4, 85748 Garching, Germany}
\address{$^3$ Nanosystems Initiative Munich (NIM), Schellingstra\ss e 4, 80339 M\"{u}nchen, Germany}
\address{$^4$ Institute for Advanced Study (IAS), Technische Universit\"at M\"unchen, Lichtenbergstra\ss e 2a, 85748 Garching, Germany}
\address{$^5$ Center for NanoScience {\it CeNS}, Geschwister-Scholl-Platz 1, 80539 M\"{u}nchen, Germany}

\ead{$^*$ hubert.krenner@physik.uni-augsburg.de}

\begin{abstract}

The excitonic occupancy state of a single, nanowire-based, heterostructure quantum dot is dynamically programmed by a surface acoustic wave. The quantum dot is formed by an interface or thickness fluctuation of a $\rm GaAs$ QW embedded in a $\rm AlGaAs$ shell of a $\rm GaAs-AlGaAs$ core-shell nanowire. As we tune the time at which carriers are photogenerated during the acoustic cycle, we find pronounced intensity oscillations of neutral and negatively charged excitons. At high acoustic power levels these oscillations become anticorrelated which enables direct acoustic programming of the dot's charge configuration, emission intensity and emission wavelength. Numerical simulations confirm that the observed modulations arise from acoustically controlled modulations of the electron and electron-hole-pair concentrations at the position of the quantum dot.

\end{abstract}

\vspace{2pc}
\vspace{2pc}
\noindent\submitto{\JPD}

\maketitle

\section{Introduction}

The dynamic control of spin and charge excitations in semiconductor nanosystems is of paramount importance for applications in quantum-optoelectronic devices. By confining the motion of carriers in one, two or all three spatial dimensions, quantum wells (QWs), quantum wires (QWRs) and Quantum Dots (QDs) have been realized on planar substrates and studied in great detail over the past decades. In the field of optically active QWs and QDs several key experiments have been performed, including for example the isolation of individual "natural", interface fluctuation QDs in a disordered QW \cite{Brunner:92}, bright, electrically driven single photon emission \cite{Yuan:2002} or the creation of "artificial" molecules with tunable bonds \cite{Krenner:06}. A central goal in the active field of fundamental and applied research are semiconductor nanowires (NW). Here, the transfer of these concepts on this one-dimensional platform using axial \cite{Borgstrom:05a,Tribu:08a} or radial approaches \cite{Qian2005,Morral:08b,Fickenscher2013} are of crucial importance. For axial NW based QDs key experiments including ultra-bright single photon emission \cite{Reimer2012}, quasi-static charge state control \cite{Kouwen:10a} and optical initialization of spin states \cite{Weert:09a} clearly underpinned the large potential of these systems.\\
Radio frequency surface acoustic waves (SAWs) represent a particularly attractive and powerful tool to probe and dynamically control charge excitations in semiconductor heterostructure including Quantum Hall systems \cite{Wixforth1986,Willett1990,Kukushkin:09}, charge transport in one- and two-dimensional electron channels \cite{Talyanskii1997,Rotter:99a}, transport of charges \cite{Rocke:97,Alsina:01,Garcia2004}, spins \cite{Sogawa:01a} or dipolar excitons \cite{Rudolph:07} and precisely timed carrier injection into QDs for low-jitter single photon emission \cite{Wiele:98,Boedefeld:06,Couto:09,Voelk:12}. Recently, these concepts have been transferred to intrinsic nanowires (NWs) \cite{Kinzel:11} and nanotubes \cite{Regler2013} and NWs containing complex radial and axial heterostructures \cite{Hernandez:12,Buyukkose2014,Weiss2014a}.\\

In this contribution we demonstrate that the excitonic occupancy state of a single NW-based heterostructure QD can be dynamically programmed by a SAW. The QD studied is formed by an interface or thickness fluctuation of a thin \emph{radial} $\rm GaAs$ QW embedded in a $\rm AlGaAs$ shell of a $\rm GaAs$ NW. As we tune the time at which carriers are photogenerated over the SAW cycle we find pronounced intensity oscillations of neutral and negatively charged excitons confined in the QD. At high acoustic power levels these oscillations become anticorrelated which enables direct acoustic programming of the QD charge configuration, emission intensity and emission wavelength. We compare the observed emission characteristics of the QD to numerical calculations of the spatio-temporal carrier dynamics in the QW induced by the SAW and find that the observed modulations arise from enhancements of the electron ($e$) and electron-hole- ($e$-$h$-) pair concentrations at the position of the QD.

\section{Sample and optical characterization}

\begin{figure}[htb]
	\begin{center}
		\includegraphics[width=0.99\textwidth]{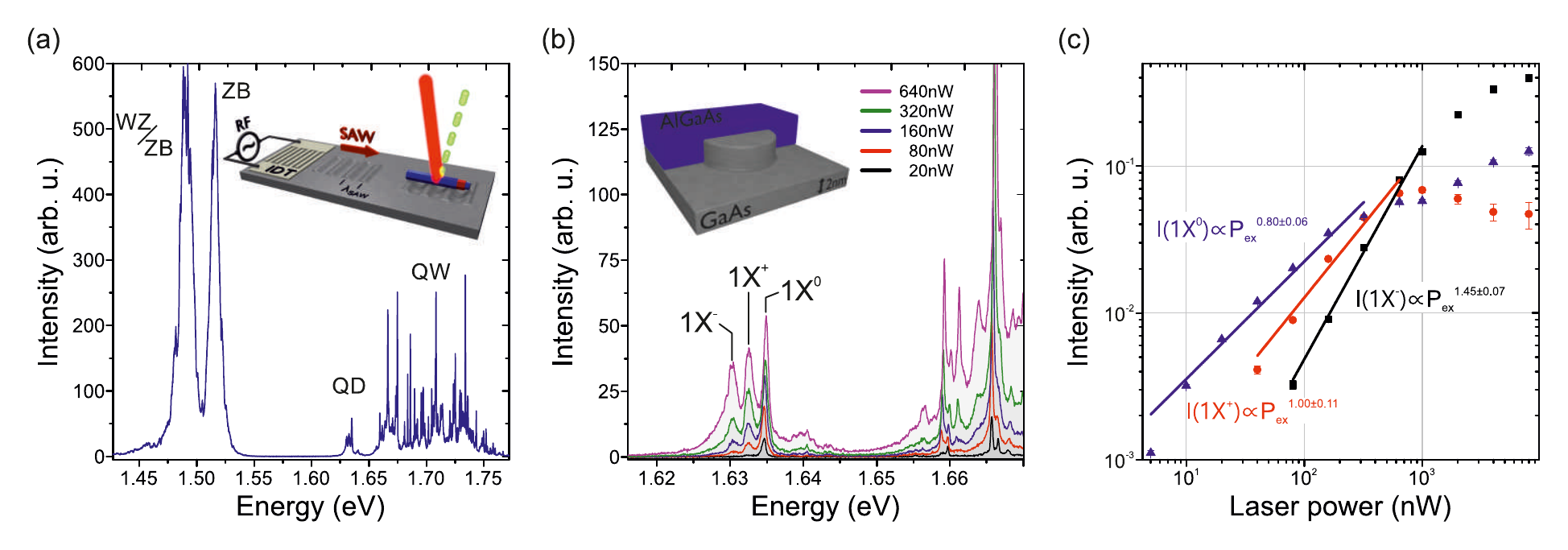}
		\caption{-- {\bf Optical characterization} -- (a) Overview spectrum exhibiting signatures of pure ZB and mixed WZ/ZB phases. The emission of a single interface fluctuation QD is detected on the \textit{low energy part} of the emission band of the 2\,nm QW. Inset: Schematic of the experimental setup. (b) PL of a single interface fluctuation QD (schematic) as a function of excitation power showing the emergence of charged excitons ($1X^{+}$, $1X^{-}$ in addition to the neutral exciton $1X^{0}$. (c) Extracted emission intensities of $1X^{0}$ (blue triangles), $1X^{+}$ (red circles) and $1X^{-}$ (black squares) as a function of the optical pump power.}
		\label{fig:1}
	\end{center}
\end{figure}

The NWs studied here were grown by solid-source molecular beam epitaxy (MBE) in a self-catalyzed growth process on a silicon substrate.\cite{Rudolph2013} Under the selected growth conditions, these NWs exhibit predominant zincblende (ZB) phase but with twin defects and short segments of wurtzite (WZ) crystal structure. The average length of the NWs was $l_{\rm NW}=10\,\mu{\rm m}$. In the radial direction the as-grown NWs consist of a $\sim60$\,nm diameter GaAs core capped by a 60 nm thick $\rm{Al_{0.3}Ga_{0.7}As}$ shell. In the center region of this shell we included a 2\,nm thick radial GaAs quantum well (QW).  Finally the NWs were passivated by a 5\,nm GaAs layer to prevent oxidation. For SAW experiments we mechanically removed NWs from the Si substrate and transferred them from suspension onto a YZ-cut $\rm LiNbO_3$ SAW-chip with lithographically defined interdigital transducers (IDTs). An RF signal applied to the IDT excites a Rayleigh-type SAW which propagates at the speed of sound, $c_{\rm SAW}=3488\,{\rm m/s}$, along the Z-direction of the $\rm LiNbO_3$ substrate. The design of the IDTs determines the SAWs wavelength to be $\lambda_{\rm SAW}=18\mu\mathrm{m}$, which corresponds to a resonance frequency of $f_{\rm SAW}=194 \, \mathrm{MHz}$ and a SAW period of $T_{\rm SAW} = 5.15 \,{\rm ns}$. After transfer, we identified NWs with their $(111)$ growth axis aligned along the SAW's propagation direction.\cite{Kinzel:11}. A schematic of our sample structure is presented as an inset of Fig. \ref{fig:1} (a).  We study the optical emission of these NWs by low-temperature $(T=10\,{\rm K})$ micro-photoluminescence ($\mu$-PL). Electron-hole pairs are photogenerated by an externally triggered, pulsed diode laser $(E_{\mathrm{laser}}=1.88\,{\rm eV})$ which was focused by a 50$\times$ microscope objective to a $\sim2\,\mu$m diameter spot. The NW emission was collected, dispersed by a 0.5 m grating monochromator and detected time integrated by a liquid $\rm N_2$ cooled Si-charge coupled device (CCD) camera.\\

In Figure \ref{fig:1} (a) we present an overview PL spectrum recorded in the center of a typical NW for an intermediate optical pump power. In this spectrum we identify the emission of the NW core at the lowest photon energies. It consists of a characteristic double peak structure.\cite{Spirkoska:09} The high energy peak at $E_{\rm ZB}=1.515\,{\rm eV}$ arises from exciton recombination in ZB regions of the NW core. The low energy peak at $E_{\rm ZB/WZ}=1.488\,{\rm eV}$ stems from recombination of spatially indirect excitons of reduced binding energy for which $e$'s and $h$'s are localized in ZB and WZ segments, respectively. The emission of the 2\,nm QW extends over a $\sim 100\,{\rm meV}$ wide energy band centered at $E_{QW}=1.7\,{\rm eV}$. {This emission is not observed for a reference sample without a radial QW, which excludes other types of emitter systems\cite{Rudolph2013,Heiss2013a,Weiss2014a} as its origin.} The large spread of the emission energy is from the (i) the small nominal thickness and (ii) the fact that growth occurs on the [110]-type facet of a NW. The QW emission peak consists of a series of sharp emission lines. These characteristics are indicative of a disordered QW with pronounced exciton localization which for planar QWs occurs due to local QW thickness fluctuations \cite{Hess:1994}. In addition to this established { and dominant type-I} contribution { to the QD confinement}, further localization can occur in our NW-based QWs due to a transfer of the ZB and WZ crystal phases from the core to the radial shell \cite{Algra2011,Rudolph2013b}{. Thus axial type-II heterostructures could also contribute to the total confinement potential. This weaker contribution would only be effective along the NW growth axis and not along the NW circumference and on its own is not sufficient to achieve the observed full three-dimensional confinement.} The such formed 'natural', interface fluctuation QDs have been studied over the past in type-I, direct bandgap GaAs/AlGaAs QWs \cite{Brunner:94a,Gammon1996,Brunner:94b,Bracker:05} as well as in type-II, indirect bandgap GaAs/AlAs QWs \cite{Zrenner:94}. The superior optical quality of this type of QDs manifests itself in long exciton coherence times \cite{Li:03} and clean single photon emission \cite{Hours2003}.\\

In the spectrum shown in Fig. \ref{fig:1} (a) the emission signal of an individual QD is well isolated on the low energy side of the QW emission band. Its energy levels are confined $\sim 20\,{\rm meV}$ below the two-dimensional QW band in which the SAW can induce spatio-temporal carrier dynamics. We want to note at this point, that this confinement energy is significantly smaller than that for self-assembled QDs and their two-dimensional wetting layer. The close examination in Fig. \ref{fig:1} (b) shows that it consists of three sharp emission lines. We attribute the dominant emission line at $E_{1X^0}=1.6349\,{\rm eV}$ to recombination of the neutral exciton, $1X^0=1e+1h$ and the two weaker signatures at $E_{1X^+}=1.6326\,{\rm eV}$ and $E_{1X^-}=1.6302\,{\rm eV}$ to the positive, $1X^+=1e+2h$, and negative trion, $1X^-=2e+1h$, respectively. At this point our assignment is based on (i) the observed renormalization energies consistent with the values reported for IFQDs in planar QWs \cite{Bracker2005} and (ii) the optical excitation power dependence of the emission intensities.\cite{Brunner:94a} The latter is plotted in double-logarithmic representation in Fig. \ref{fig:1} (c) to identify the underlying power law dependencies. Clearly, the line assigned to $1X^0$ and $1X^+$ exhibits the smallest slope of 0.8 and 1.0, consistent with a single excitonic nature. In contrast, $1X^-$ exhibits a larger slope of 1.45 which could be indicative for a charged exciton or the biexciton. { Our excitation power dependent spectroscopy does not show signatures of higher confined shells. The observed broadening most likely arises from interactions of the confined exciton species with free or localized carriers in the surrounding QW. Thus, higher charged exciton species, which require carriers populating higher shells can be excluded. Moreover, the QD emissions shows a weak ($<1\,{\rm meV}$) \emph{blueshift} with increasing excitation power which points towards the presence of an additional contribution of a type-II band alignment due to polytypism. Finally, we want to note at this point, that our} SAW experiments presented later will provide clear evidence for our attribution of this emission line to $1X^-$.

\section{Dynamic spectral modulation and SAW phase calibration}

\begin{figure}[htb]
	\begin{center}
		\includegraphics[width=0.99\textwidth]{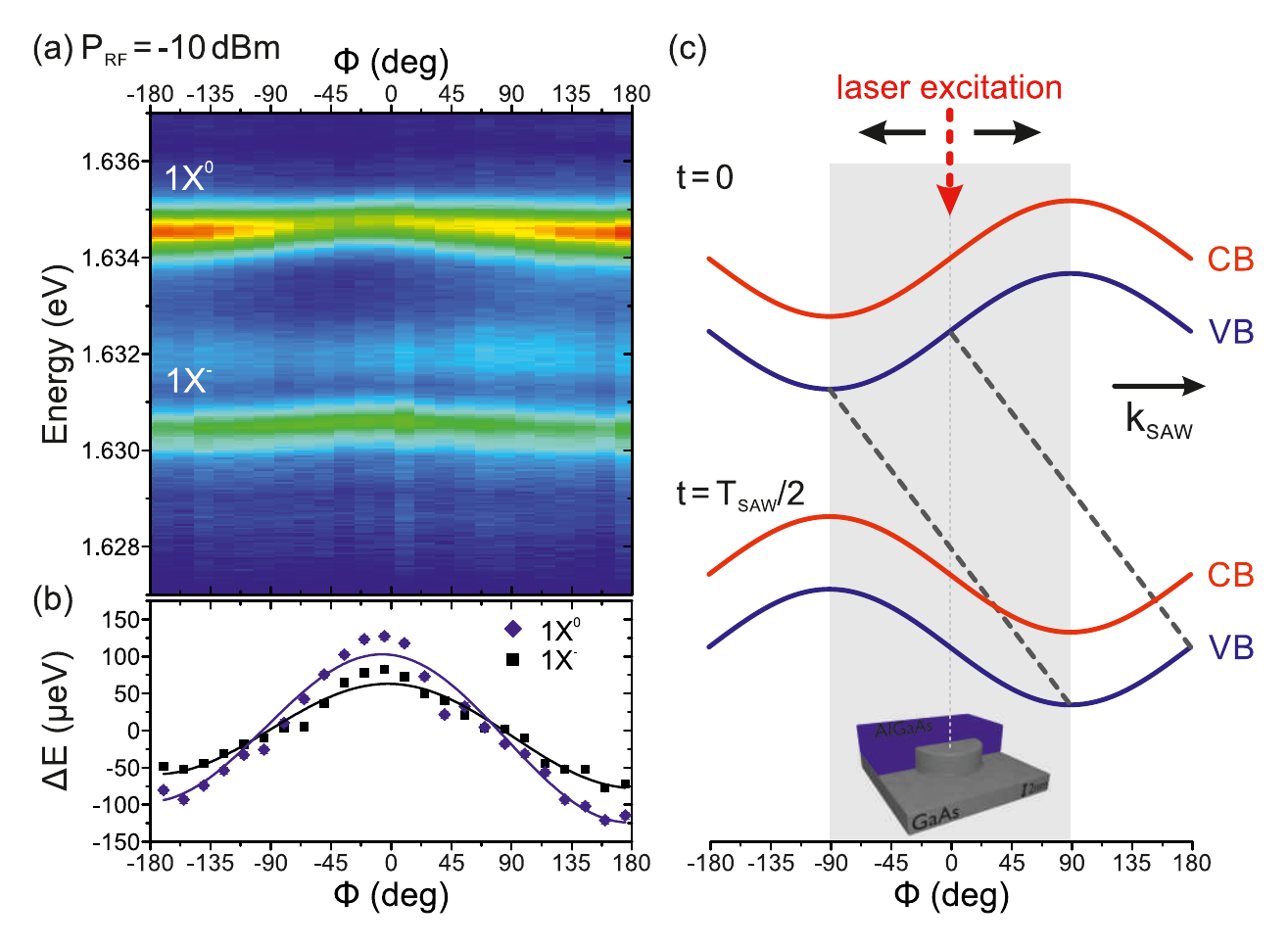}
		\caption{-- {\bf Spectral modulation, phase calibration and temporal evolution of the band edge modulation} -- (a) Stroboscopic PL spectra at low acoustic amplitudes $(P_{\rm RF} = -10 \,{\rm dBm})$ as a function of $\phi$ and photon energy with the normalized intensity encoded in false color (blue: low intensity, red: maximum intensity). b) Extracted spectral shifts of $1X^{0}$ (diamonds) and $1X^{-}$ (squares). The solid lines are fits to the experimental data to calibrate $\phi$.(c) Schematic of the temporal evolution of the SAW-induced bandedge modulation. The grey shaded region indicates the range of $\phi$ corresponding to the NW length in our experiment.}
		\label{fig:2}
	\end{center}
\end{figure}

We now turn to the manipulation of the QD's emission by the dynamic strain and electric fields of a SAW. To assess the dynamic SAW-driven modulation of the optical emission we employ an established stroboscopic excitation scheme combined with time-integrated multichannel detection \cite{Voelk:11a,Fuhrmann:11}. { We employ weak optical pumping which is crucial for these experiments for three reasons: (i) The resulting low charge carrier densities in the QW and NW core do not significantly screen the SAW-induced electric fields driving the spatio-temporal carrier dynamics. The maximum number of (ii) excitons and (iii) excess $e$'s captured by the QD is limited to 1.} In Fig. \ref{fig:2} (a) we show stroboscopic PL data of the above characterized QD controlled by a SAW excited by applying a radio frequency signal of power $P_{\rm RF} = -10 \,{\rm dBm}$ to the IDT. The emission intensity is color coded and plotted as a function of the photon energy (vertical axis) and stroboscopic phase $\phi$ (horizontal axis). Both emission lines observed in the data, $1X^0$ and $1X^-$, exhibit clear sinusoidal spectral modulations due to the dynamic strain field of the SAW. This strain field and its corresponding hydrostatic pressure $(p)$ tunes the QD emission due to deformation potential coupling \cite{Pollak1968}. At such low acoustic amplitudes the contribution of SAW-induced lateral electric fields are weak \cite{Kaniber:11a} and the observed spectral shift is dominated by the strain contribution. \cite{Gell:08,Weiss2014a} The extracted relative spectral tuning $\Delta E_{strain}$ of $1X^0$ and $1X^-$ are plotted as a function of $\phi$ in Fig. \ref{fig:2} (b). Clearly, the modulations of both emission lines are in phase and exhibit similar amplitudes. This is expected because the underlying deformation potential coupling modulates the effective bandgap with a weaker contribution arising from a perturbation of the single particle wavefunctions. \cite{Joens:11} The observed spectral shifts correspond to $0.5<p<0.8\,{\rm MPa}$ \cite{Pollak1968}. Taken together, the maximum (minimum) emission energies are observed for the value of $\phi$ at which $p$ is maximum positive (negative). For the crystal cut of our $\rm LiNbO_3$ substrate we can derive the $\phi$-dependence of the SAW-induced modulation of the conduction (CB) and valence band (VB) within the GaAs NW using finite element modeling (FEM) \cite{Weiss2014a} which is shown in Fig. \ref{fig:2}(c) for times $t=0$ (i.e. time of photoexcitation) and $t=T_{\rm SAW}/2$ at which the band modulation is translated spatially due to the propagation of the SAW. Using this calibration reveals that the weak suppression (enhancement) of the $1X^0$ $(1X^-)$ emission occurs for stroboscopic excitation in between the stable $(\phi=-90^{\rm o})$ and unstable $(\phi=+90^{\rm o})$ points for $e$'s in the CB and for which the $e$-drift is anti-parallel to the SAW propagation and wavevector $k_{\rm SAW}$. Thus, as the SAW propagates, $e$'s are effectively \emph{transported back} to the position of the QD and increase the net $e$-density which in turn \emph{reduces} the probability of the QD being occupied by $1X^0$. Therefore, the SAW effectively induces a anti-correlated rocking motion for $e$'s and $h$'s which is weak at the applied small band edge modulation. This picture does not take into account the \emph{finite} length of the NW. Before presenting detailed experimental and numerical results of the SAW control of the QD's occupancy state, we want to point out, that we have to take into account the fact that $l_{\rm NW}\simeq\lambda_{\rm SAW}/2$. This boundary condition limits the spatio-temporal carrier dynamics to a sub-$\lambda_{\rm SAW}$ lengthscale, which is in strong contrast to previous experiments performed on planar self-assembled QDs and Quantum Posts \cite{Voelk:10b,Voelk:11a,Schulein2013}. The two ends of the NW effectively represent efficient recombination sites for carriers and non-radiatively remove these from the system. We convert the spatial coordinate along the NW axis $x$ to $\phi$ and take into account (i) QD position in the center of the NW $(x=0)$ and (ii) $l_{\rm NW}\simeq\lambda_{\rm SAW}/2$. The latter condition implies that, for a fixed value of $\phi=\phi_0$, carrier can be transferred without loss by $\Delta x \sim  \lambda_{\rm SAW}/4$ away from the point of photogeneration, i.e. the position of the QD in the center of the NW. This restriction is indicated by the grey shaded area in Figure \ref{fig:2} (c).

\section{Acoustic control of QD occupancy state}

In this section we investigate the SAW-regulated occupancy state control in detailed stroboscopic PL experiments and numerical simulations of the SAW-driven spatio-temporal carrier dynamics.\\

\begin{figure}[htb]
	\begin{center}
		\includegraphics[width=0.99\textwidth]{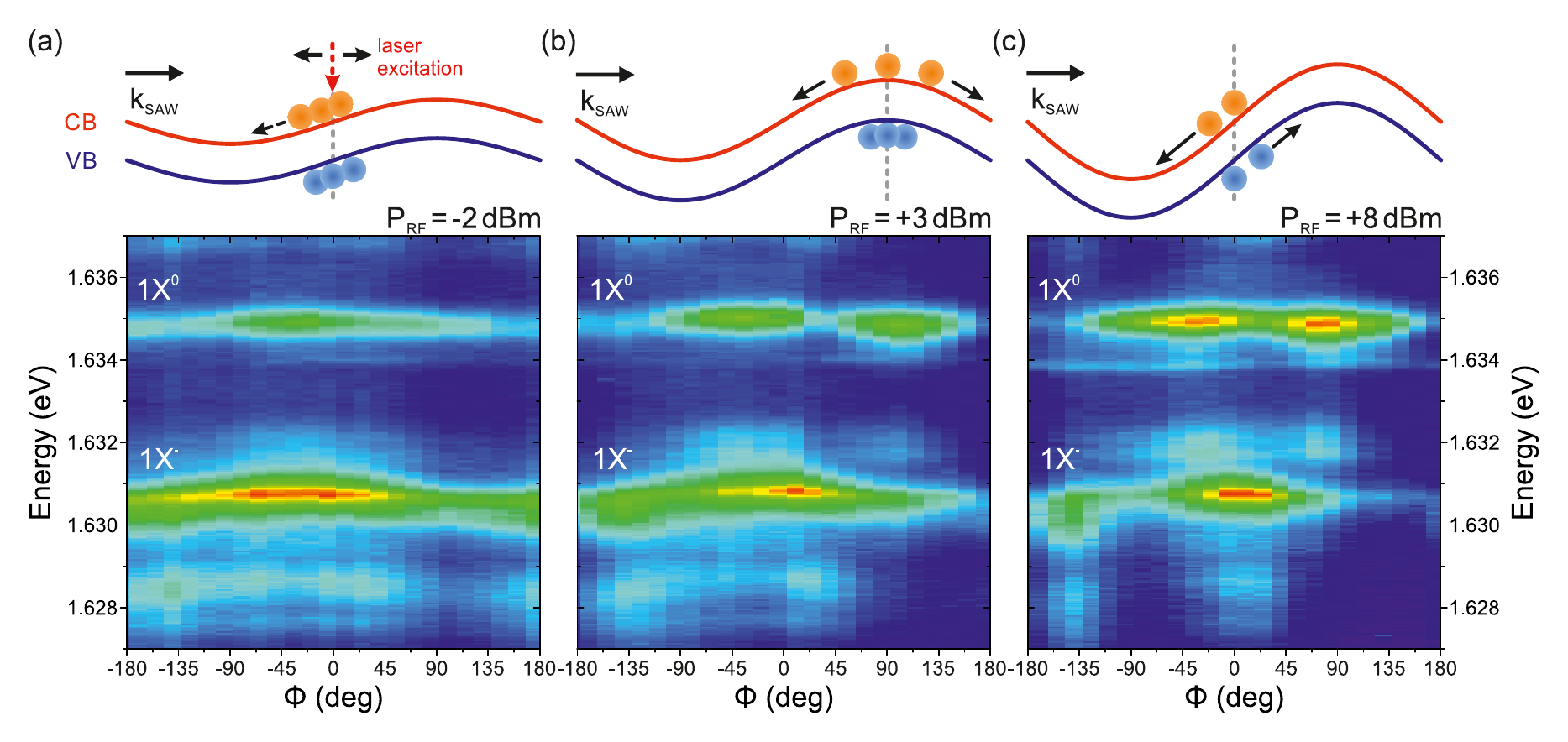}
		\caption{-- {\bf SAW control of QD occupancy state} -- Stroboscopic PL spectra in falsecolor representation for (a) $P{\rm _{RF}= -2\,dBm}$, (b) $\rm +3\,dBm$,  (c) $\rm +8\,dBm$ and the corresponding bandedge modulations as schematics in the upper part of each panel: Maximum $1X^-$ intensity is detected for $\phi\leq0^{\rm o}$, schematics in (a+c). The additional maximum of $1X^0$ develops for $\phi\simeq 90^{\rm 0}$, schematics in (b).}
		\label{fig:3}
	\end{center}
\end{figure}

We begin by presenting stroboscopic PL data recorded from the same QD for $P{\rm _{RF}= -2\,dBm,\, +3\,dBm,\, +8\,dBm}$ in Fig. \ref{fig:3}(a-c). The data are plotted in the same representation as in Fig. \ref{fig:2} (a) and the corresponding CB and VB modulations are sketched schematically in the upper parts of each panel. As we increase $P{\rm _{RF}}$, the intensity modulations of the dominant $1X^0$ and $1X^-$ emissions changes dramatically. Most notably, the $P{\rm _{RF}= +8\,dBm}$, the $\phi$-dependence of the $1X^0$ and $1X^-$ emission intensities are clearly anticorrelated. This anticorrelation provides a direct route to effectively program the occupancy state of the QD simply by tuning $\phi$ at the RF signal generator used to excite the SAW. \\
The first striking difference between this data and the data in Fig. \ref{fig:1} is the pronounced increase of the $1X^-$ emission, which is more intense than that of $1X^0$. { This observation confirms an increase of probability to statistically capture and additional $e$ due to the SAW-induced carrier dynamics analogous to our recent experiments preformed on InGaAs QDs on planar substrates \cite{Schulein2013}. We want to note that the QDs studied in Ref. \cite{Schulein2013} confine a larger number of charges compared to the QD studied here, no indications of higher charged excitons species are observed under the applied weak optical excitations. This provides further evidence for the assignment of the different emission lines in the experiments presented here. Moreover, the} higher intensity of $1X^-$ provides direct evidence that this emission line does not arise from $2X^0$ recombination which should exhibit at most the same intensity as $1X^0$ under the applied pulsed optical excitation scheme. We continue by addressing a second particularly peculiar feature when comparing the data at these three acoustic amplitudes. It is the development of a global minimum of the QD emission at $\phi=\pm180^{\rm o}$. At this local stroboscopic phase the drift of electrons in the SAW-induced electric field is aligned with the SAW propagation. Thus, both effects add up and result in an efficient depletion of the $e$-concentration at the position of the QD and subsequent loss of this carrier species. Taking into account that $e$'s exhibit a larger transport mobility than $h$'s, the global minimum of the QD occupancy and emission is indeed expected at this local phase confirming our $\phi$-calibration. As we increase $P{\rm _{RF}}$, two prominent effects are observed: (i) The $1X^0$ emission develops two pronounced maxima at $\phi=-45^{\rm o}$ and $\phi=+90^{\rm o}$. (ii) The range of $\phi$ over which $1X^-$ is observed focuses to a $\Delta \phi=90^{\rm o}$ wide range centered around $\phi=0$. The increase of $1X^-$ can be qualitatively understood by an acoustically regulated return of $e$'s to the position of the QD as described in the previous section and shown in the schematics of Figures \ref{fig:2} (c) and \ref{fig:3} (a). Furthermore, at high acoustic amplitudes, the first step of this process, the field-driven transfer of $e$'s to their stable CB minimum dominates. When $\phi$ is tuned to positive values, this minimum shifts outside of the NW. Thus, non-radiative loss of $e$'s is enhanced for $45^{\rm o}<\phi<90^{\rm o}$. The upper limit of $\phi=90^{\rm o}$ (cf. schematic in Fig. \ref{fig:3} (b)) corresponds to photogeneration of $e$'s at an instable point, the maximum of the CB modulation, while $h's$ are effectively confined at the stable VB maximum. The combination of these effects enhance the $e$-$h$-pair, exciton density at this stroboscopic phase and in turn favours pair-wise capture. Similar reasoning can be applied for $\phi=-90^{\rm o}$ at which the exciton formation is expected to be enhanced further since the transfer of $h$'s toward their stable point in the VB is comparably slow due to their reduced transport mobility. This effect leads to the observed formation of $1X^0$ in the dot. For $0^{\rm o}>\phi>-90^{\rm o}$ the removal of $h$'s along the SAW propagation direction breaks down (cf. schematic in Figure \ref{fig:3} (c)) giving rise to a net increase of the $h$-density and in turn a reduced probability for forming $1X^-$. Based on these so far qualitative arguments we attribute the preferential formation of $1X^0$ to an enhanced (neutral) exciton concentration at the position of the QD and the preferential formation of $1X^-$ to that of $e$'s. Before presenting numerical simulations of the underlying SAW-driven carrier dynamics we want to point out that the observed intensity modulations and their dependence on $P_{\rm RF}$ is similar to that observed for planar self-assembled QDs \cite{Schulein2013}.{ T}he data reported here clearly reproduces the expected $\phi$-dependence of the \emph{horizontal} electric field component which is aligned with the NW axis as it is also the case for planar QDs and quantum posts. The combination of these observations provides final evidence for our assignment of the observed emission lines as arising from exciton recombination in an interface fluctuation QD.\\

\begin{figure}[htb]
	\begin{center}
		\includegraphics[width=0.99\textwidth]{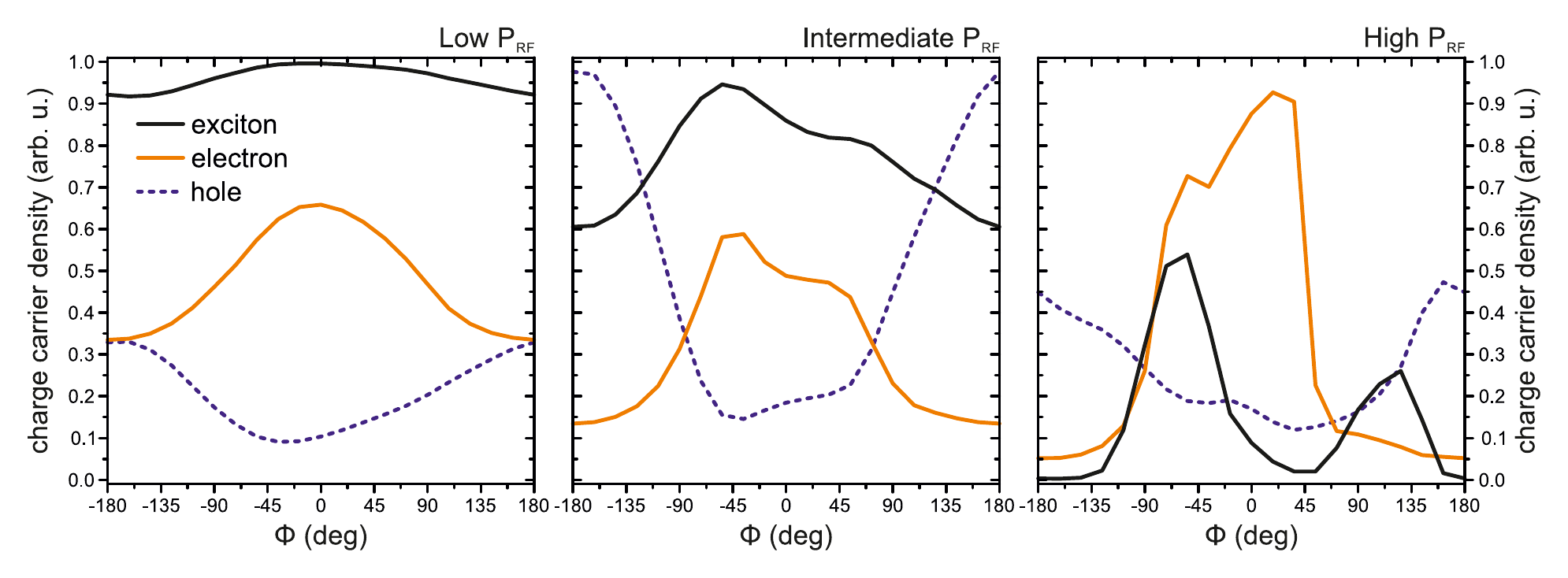}
		\caption{-- {\bf Simulation of SAW-controlled electron and exciton densities} -- Calculated time-integrated exciton (black){, electron (orange) and hole (dashed, blue) } densities (in arbitrary units) at the position of the QD as a function of $\phi$ for low (a), intermediate (b) and high (c) acoustic amplitudes.}
		\label{fig:4}
	\end{center}
\end{figure}
To confirm that this qualitative picture indeed agrees with the underlying spatio-temporal carrier dynamics, we performed numerical solutions of the semi-classical drift and diffusion processes in the SAW-modulated bandstructure. In contrast to { a QW which infinitely extends in the plane as }assumed in previous reports \cite{Garcia2004,Schuelein:12a}, we take into account the geometry of the NW studied in our experiments. We account for its finite length $(l_{\rm NW}\simeq\lambda_{\rm SAW}/2)$ and the non-radiative loss of carriers at the open facets at the NW ends by removing carriers arriving at position $|x|>l_{\rm NW}/2=\lambda_{\rm SAW}/4$ from the center of the NW $(x=0)$. In particular we integrated the calculated  $e${, $h$} and exciton ($e$-$h$-pair) densities at the QD position over time as a function of $\phi$. The $\phi$-dependence of these time-integrated densities are plotted in Figure \ref{fig:4} for low (a), intermediate (b) and high acoustic powers (c), as black and orange lines, respectively. For low acoustic powers ($P{\rm _{RF}}$) the exciton density remains high over the entire range of $\phi$ which is consistent with the overall modulation observed for $1X^0$ and also $1X^-$. In contrast, the calculated $e$ density exhibits a clear and broad maximum centered at $\phi\sim - 15^{\rm o}$. This maximum nicely overlaps with that of $1X^-$ observed in the experimental data in Figure \ref{fig:3} (a). As we increase the acoustic amplitude in our simulations (cf. Figure \ref{fig:4} (b+c))  the exciton density develops a clear double peak structure in good agreement with the experimentally observed maxima of the $1X^0$ modulation. Furthermore, also the calculated $e$-density modulation sharpens in $\phi$ which reproduces the modulations of $1X^0$ in the experimental data. In addition, an increase of the $e$ density with respect to that of excitons with increasing acoustic power is clearly resolved in this simulation data. This calculated behavior is in good agreement with the experimentally observed increase of the $1X^-$ intensity with respect to that of $1X^0$ (cf. Figures \ref{fig:2} and \ref{fig:3}).{ Furthermore, the $\phi$-dependency of $h$ density is anticorrelated to that of $e$ providing further evidence for the the preferential formation of \emph{negatively} charged excitons.} \\
{ The emission lines similar to those reported by Heiss \emph{et al.} in \cite{Heiss2013a} exhibit also a characteristic SAW-response \cite{Weiss2014a}. This, however, features characteristic differences to that reported here  since both the emission center and the driving mechanism are of different nature:
(i) In Ref. \cite{Weiss2014a} we identify quantum tunneling as the underlying mechanism and the overall $\phi$-dependence of the remains constant within the experimental resolution as $P_{\rm RF}$ is tuned. (ii) This fixed $\phi$-dependence follows the oscillation of the vertical electric field component of the SAW which dynamically modulates the tunneling probability of carriers out of randomly distributed, defect-related emission centers localized in the $\rm Al_{0.7}Ga_{0.3}As$ shell to a continuum of states. Both effects are in strong contrast to the data presented here and similar experiments performed for planar QD \cite{Schulein2013} and quantum post systems \cite{Voelk:11a}. Moreover, the calculated tunneling rates using WKB theory indicate that for the radial layer sequence chosen here, \emph{any type of emission center} localized in the radial shell is expected to be depopulated non-radiatively due to quantum tunneling as discussed in Ref. \cite{Weiss2014a} and its accompanying supplementary information.}

\section{Conclusions and outlook}
In summary, we have demonstrated that the occupancy state of a NW-based heterostructure QD can be programmed using SAW-regulated carrier injection under strictly stroboscopic optical excitation. In our experiments we observe characteristic intensity oscillation of $1X^0$ and $1X^-$ as we tune the stroboscopic excitation phase $\phi$. The modulation show a characteristic dependence on the acoustic amplitude. These arise from a modifications of the time-integrated exciton and $e$ densities at the position of the QD as confirmed by numerical simulations of the SAW-driven spatio-temporal carrier dynamics. Our results clearly demonstrate that concepts developed for planar semiconductor heterostructures can be readily transferred to a NW-platform. Our experiments have been performed on \emph{radial} heterostructure and could be directly applied to axial heterostructures. For this type of architecture carrier extraction schemes, as recently demonstrated for NW-based defect-related emission centers \cite{Weiss2014a} could be combined with the here reported injection principle to realize SAW-mediated single carrier transfer between distant \emph{optically active} QDs on a single NW \cite{Hermelin:11,McNeil:11}. { To implement quantum communication protocols based on the spin degree of freedom, these states have to be initialized optically. For NWs this has been first demonstrated both for vertically oriented, standing \cite{Weert:09a} NWs. Recently, the polarizing effect of the NW has been overcome for a lying geometries  \cite{Bulgarini2012a} fully compatible with our SAW technique.}

\section*{Acknowledgements}
This work was supported by the Deutsche Forschungsgemeinschaft (DFG) via Sonderforschungsbereich SFB631 (Projects B1 and B5) and the Emmy Noether Program (KR 3790/2-1) and by the European Union via SOLID and the FP7 Marie-Curie Reintegration Grant.

\section*{References}

\begin{thebibliography}{10}
\expandafter\ifx\csname url\endcsname\relax
  \def\url#1{{\tt #1}}\fi
\expandafter\ifx\csname urlprefix\endcsname\relax\def\urlprefix{URL }\fi
\providecommand{\eprint}[2][]{\url{#2}}

\bibitem{Brunner:92}
Brunner K, Bockelmann U, Abstreiter G, Walther M, B\"{o}hm G, Tr\"{a}nkle G and
  Weimann G 1992 {\em Physical Review Letters\/} {\bf 69} 3216--3219

\bibitem{Yuan:2002}
Yuan Z, Kardynal B~E, Stevenson R~M, Shields A~J, Lobo C~J, Cooper K, Beattie
  N~S, Ritchie D~A and Pepper M 2002 {\em Science\/} {\bf 295} 102--105
  

\bibitem{Krenner:06}
Krenner H~J, Clark E~C, Nakaoka T, Bichler M, Scheurer C, Abstreiter G and
  Finley J~J 2006 {\em Physical Review Letters\/} {\bf 97} 076403 

\bibitem{Borgstrom:05a}
Borgstr\"{o}m M~T, Zwiller V, M\"{u}ller E and Imamoglu A 2005 {\em Nano
  Letters\/} {\bf 5} 1439--1443 

\bibitem{Tribu:08a}
Tribu A, Sallen G, Aichele T, Andr\'{e} R, Poizat J~P, Bougerol C, Tatarenko S
  and Kheng K 2008 {\em Nano Letters\/} {\bf 8} 4326--9 

\bibitem{Qian2005}
Qian F, Gradecak S, Li Y, Wen C~Y and Lieber C~M 2005 {\em Nano Letters\/} {\bf
  5} 2287--91 

\bibitem{Morral:08b}
{Fontcuberta i Morral} A, Spirkoska D, Arbiol J, Heigoldt M, {Ramon Morante} J
  and Abstreiter G 2008 {\em Small\/}
  {\bf 4} 899--903 

\bibitem{Fickenscher2013}
Fickenscher M, Shi T, Jackson H~E, Smith L~M, Yarrison-Rice J~M, Zheng C,
  Miller P, Etheridge J, Wong B~M, Gao Q, Deshpande S, Tan H~H and Jagadish C
  2013 {\em Nano Letters\/} {\bf 13} 1016--22 

\bibitem{Reimer2012}
Reimer M~E, Bulgarini G, Akopian N, Hocevar M, Bavinck M~B, Verheijen M~A,
  Bakkers E~P~A~M, Kouwenhoven L~P and Zwiller V 2012 {\em Nature
  Communications\/} {\bf 3} 737 

\bibitem{Kouwen:10a}
van Kouwen M~P, Reimer M~E, Hidma A~W, van Weert M~H~M, Algra R~E, Bakkers
  E~P~A~M, Kouwenhoven L~P and Zwiller V 2010 {\em Nano Letters\/} {\bf 10}
  1817--22 

\bibitem{Weert:09a}
van Weert M~H~M, Akopian N, Perinetti U, van Kouwen M~P, Algra R~E, Verheijen
  M~A, Bakkers E~P~A~M, Kouwenhoven L~P and Zwiller V 2009 {\em Nano Letters\/}
  {\bf 9} 1989--1993

\bibitem{Wixforth1986}
Wixforth A, Kotthaus J and Weimann G 1986 {\em Physical Review Letters\/} {\bf
  56} 2104--2106 

\bibitem{Willett1990}
Willett R, Paalanen M, Ruel R, West K, Pfeiffer L and Bishop D 1990 {\em
  Physical Review Letters\/} {\bf 65} 112--115 

\bibitem{Kukushkin:09}
Kukushkin I~V, Smet J~H, Scarola V~W, Umansky V and von Klitzing K 2009 {\em
  Science\/} {\bf 324} 1044--1047

\bibitem{Talyanskii1997}
Talyanskii V, Shilton J, Pepper M, Smith C, Ford C, Linfield E, Ritchie D and
  Jones G 1997 {\em Physical Review B\/} {\bf 56} 15180--15184 

\bibitem{Rotter:99a}
Rotter M, Kalameitsev A~V, Govorov A~O, Ruile W and Wixforth A 1999 {\em
  Physical Review Letters\/} {\bf 82} 2171--2174

\bibitem{Rocke:97}
Rocke C, Zimmermann S, Wixforth A, Kotthaus J~P, B\"{o}hm G and Weimann G 1997
  {\em Physical Review Letters\/} {\bf 78} 4099--4102

\bibitem{Alsina:01}
Alsina F, Santos P~V, Hey R, Garc\'{\i}a-Crist\'{o}bal A
  and Cantarero A 2001 {\em Physical Review B\/} {\bf 64} 041304(R)

\bibitem{Garcia2004}
Garc\'{\i}a-Crist\'{o}bal A, Cantarero A, Alsina F and Santos P~V 2004 {\em
  Physical Review B\/} {\bf 69} 205301 

\bibitem{Sogawa:01a}
Sogawa T, Santos P~V, Zhang S~K, Eshlaghi S, Wieck A~D and Ploog K~H 2001 {\em
  Physical Review Letters\/} {\bf 87} 276601

\bibitem{Rudolph:07}
Rudolph J, Hey R and Santos P~V 2007 {\em Physical Review Letters\/} {\bf 99}
  047602

\bibitem{Wiele:98}
Wiele C, Haake F, Rocke C and Wixforth A 1998 {\em Physical Review A\/} {\bf
  58} R2680--R2683 

\bibitem{Boedefeld:06}
B\"{o}defeld C, Ebbecke J, Toivonen J, Sopanen M, Lipsanen H and Wixforth A
  2006 {\em Physical Review B\/} {\bf 74} 035407 

\bibitem{Couto:09}
Couto O~D~D, Lazi\'{c} S, Iikawa F, Stotz J~A~H, Jahn U, Hey R and Santos P~V
  2009 {\em Nature Photonics\/} {\bf 3} 645--648 

\bibitem{Voelk:12}
V\"{o}lk S, Knall F, Sch\"{u}lein F~J~R, Truong T~A, Kim H, Petroff P~M,
  Wixforth A and Krenner H~J 2012 {\em Nanotechnology\/} {\bf 23} 285201 

\bibitem{Kinzel:11}
Kinzel J~B, Rudolph D, Bichler M, Abstreiter G, Finley J~J, Koblm\"{u}ller G,
  Wixforth A and Krenner H~J 2011 {\em Nano Letters\/} {\bf 11} 1512--1517 

\bibitem{Regler2013}
Regler M~E, Krenner H~J, Green A~A, Hersam M~C, Wixforth A and Hartschuh A 2013 {\em
  Chemical Physics\/} {\bf 413} 39--44 

\bibitem{Hernandez:12}
Hern\'{a}ndez-M\'{\i}nguez A, M\"{o}ller M, Breuer S, Pf\"{u}ller C, Somaschini
  C, Lazi\'{c} S, Brandt O, Garc\'{\i}a-Crist\'{o}bal A, de~Lima M~M, Cantarero
  A, Geelhaar L, Riechert H and Santos P~V 2012 {\em Nano Letters\/} {\bf 12}
  252--8 
\bibitem{Buyukkose2014}
B\"{u}y\"{u}kk\"{o}se S, Hern\'{a}ndez-M\'{\i}nguez A, Vratzov B, Somaschini C,
  Geelhaar L, Riechert H, van~der Wiel W~G and Santos P~V 2014 {\em
  Nanotechnology\/} {\bf 25} 135204 

\bibitem{Weiss2014a}
Wei\ss ~M, Kinzel J~B, Sch\"{u}lein F~J~R, Heigl M, Rudolph D, Mork\"{o}tter S,
  D\"{o}blinger M, Bichler M, Abstreiter G, Finley J~J, Koblm\"{u}ller G,
  Wixforth A and Krenner H~J 2014 {\em Nano Letters\/} {\bf 14} 2256--64 

\bibitem{Rudolph2013}
Rudolph D, Funk S, D\"{o}blinger M, Mork\"{o}tter S, Hertenberger S,
  Schweickert L, Becker J, Matich S, Bichler M, Spirkoska D, Zardo I, Finley
  J~J, Abstreiter G and Koblm\"{u}ller G 2013 {\em Nano Letters\/} {\bf 13}
  1522--7 

\bibitem{Spirkoska:09}
Spirkoska D, Arbiol J, Gustafsson A, Conesa-Boj S, Glas F, Zardo I, Heigoldt M,
  Gass M~H, Bleloch A~L, Estrade S, Kaniber M, Rossler J, Peiro F, Morante J~R,
  Abstreiter G, Samuelson L and {Fontcuberta i Morral} A 2009 {\em Physical
  Review B\/} {\bf 80} 245325 

\bibitem{Heiss2013a}
Heiss M, Fontana Y, Gustafsson A, W\"{u}st G, Magen C, O'Regan D~D, Luo J~W,
  Ketterer B, Conesa-Boj S, Kuhlmann A~V, Houel J, Russo-Averchi E, Morante
  J~R, Cantoni M, Marzari N, Arbiol J, Zunger A, Warburton R~J and {Fontcuberta
  i Morral} A 2013 {\em Nature Materials\/} {\bf 12} 439--44 

\bibitem{Hess:1994}
Hess H~F, Betzig E, Harris T~D, Pfeiffer L~N and West K~W 1994 {\em Science\/}
  {\bf 264} 1740--1745
  
\bibitem{Algra2011}
Algra R~E, Hocevar M, Verheijen M~A, Zardo I, Immink G~G~W, van Enckevort
  W~J~P, Abstreiter G, Kouwenhoven L~P, Vlieg E and Bakkers E~P~A~M 2011 {\em
  Nano Letters\/} {\bf 11} 1690--4 

\bibitem{Rudolph2013b}
Rudolph D, Schweickert L, Mork\"{o}tter S, Hanschke L, Hertenberger S, Bichler
  M, Koblm\"{u}ller G, Abstreiter G and Finley J~J 2013 {\em New Journal of
  Physics\/} {\bf 15} 113032

\bibitem{Brunner:94a}
Brunner K, Abstreiter G, B\"{o}hm G, Tr\"{a}nkle G and Weimann G 1994 {\em
  Physical Review Letters\/} {\bf 73} 1138--1141
 

\bibitem{Gammon1996}
Gammon D, Snow E, Shanabrook B, Katzer D and Park D 1996 {\em Science (New
  York, N.Y.)\/} {\bf 273} 87--90 
  
\bibitem{Brunner:94b}
Brunner K, Abstreiter G, B\"{o}hm G, Tr\"{a}nkle G and Weimann G 1994 {\em
  Applied Physics Letters\/} {\bf 64} 3320--3322
 

\bibitem{Bracker:05}
Bracker A~S, Stinaff E~A, Gammon D, Ware M~E, Tischler J~G, Shabaev A, Efros
  A~L, Park D, Gershoni D, Korenev V~L and Merkulov I~A 2005 {\em Physical
  Review Letters\/} {\bf 94} 047402
  
\bibitem{Zrenner:94}
Zrenner A, Butov L~V, Hagn M, Abstreiter G, B\"{o}hm G and Weimann G 1994 {\em
  Physical Review Letters\/} {\bf 72} 3382--3385
 
\bibitem{Li:03}
Li X, Wu Y, Steel D, Gammon D, Stievater T~H, Katzer D~S, Park D, Piermarocchi
  C and Sham L~J 2003 {\em Science\/} {\bf 301} 809--811
  
\bibitem{Hours2003}
Hours J, Varoutsis S, Gallart M, Bloch J, Robert-Philip I, Cavanna A, Abram I,
  Laruelle F and Gérard J~M 2003 {\em Applied Physics Letters\/} {\bf 82}
  2206 

\bibitem{Bracker2005}
Bracker A, Stinaff E, Gammon D, Ware M, Tischler J, Park D, Gershoni D, Filinov
  A, Bonitz M, Peeters F and Riva C 2005 {\em Physical Review B\/} {\bf 72}
  035332 
  
\bibitem{Voelk:11a}
V\"{o}lk S, Knall F, Sch\"{u}lein F~J~R, Truong T~A, Kim H, Petroff P~M,
  Wixforth A and Krenner H~J 2011 {\em Applied Physics Letters\/} {\bf 98}
  023109 

\bibitem{Fuhrmann:11}
Fuhrmann D~A, Thon S~M, Kim H, Bouwmeester D, Petroff P~M, Wixforth A and
  Krenner H~J 2011 {\em Nature Photonics\/} {\bf 5} 605--609 

\bibitem{Pollak1968}
Pollak F and Cardona M 1968 {\em Physical Review\/} {\bf 172} 816--837 

\bibitem{Kaniber:11a}
Kaniber M, Huck M~F, M\"{u}ller K, Clark E~C, Troiani F, Bichler M, Krenner H~J
  and Finley J~J 2011 {\em Nanotechnology\/} {\bf 22} 325202

\bibitem{Gell:08}
Gell J~R, Ward M~B, Young R~J, Stevenson R~M, Atkinson P, Anderson D, Jones
  G~A~C, Ritchie D~A and Shields A~J 2008 {\em Applied Physics Letters\/} {\bf
  93} 081115 

\bibitem{Joens:11}
J\"{o}ns K, Hafenbrak R, Singh R, Ding F, Plumhof J, Rastelli A, Schmidt O,
  Bester G and Michler P 2011 {\em Physical Review Letters\/} {\bf 107} 

\bibitem{Voelk:10b}
V\"{o}lk S, Sch\"{u}lein F~J~R, Knall F, Reuter D, Wieck A~D, Truong T~A, Kim
  H, Petroff P~M, Wixforth A and Krenner H~J 2010 {\em Nano Letters\/} {\bf 10}
  3399--3407 
\bibitem{Schulein2013}
Sch\"{u}lein F~J~R, M\"{u}ller K, Bichler M, Koblm\"{u}ller G, Finley J~J,
  Wixforth A and Krenner H~J 2013 {\em Physical Review B\/} {\bf 88} 085307
  

\bibitem{Schuelein:12a}
Sch\"{u}lein F~J~R, Pustiowski J, M\"{u}ller K, Bichler M, Koblm\"{u}ller G,
  Finley J~J, Wixforth A and Krenner H~J 2012 {\em JETP Letters\/} {\bf 95}
  575--580 

\bibitem{Hermelin:11}
Hermelin S, Takada S, Yamamoto M, Tarucha S, Wieck A~D, Saminadayar L,
  B\"{a}uerle C and Meunier T 2011 {\em Nature\/} {\bf 477} 435--8 

\bibitem{McNeil:11}
McNeil R~P~G, Kataoka M, Ford C~J~B, Barnes C~H~W, Anderson D, Jones G~A~C,
  Farrer I and Ritchie D~A 2011 {\em Nature\/} {\bf 477} 439--42 

\bibitem{Bulgarini2012a}
Bulgarini G, Reimer M~E and Zwiller V 2012 {\em Applied Physics Letters\/} {\bf
  101} 111112 

\end{thebibliography}

\providecommand{\newblock}{}

\end{document}